\documentclass[doublecol]{epl2} 

\usepackage{bbm}
\usepackage{color}
\usepackage{amsmath,amssymb,latexsym}

\newcommand{\Itot}{I_{\mathrm{tot}}}
\newcommand{\Iedge}{I_{\mathrm{edge}}}

\newcommand{\micron}{\mu\mathrm{m}}

\title{Network permeability changes according to a quadratic power law\\ upon removal of a single edge}
\shorttitle{Network permeability: Quadratic power law}

\author{S. Lange\inst{1} \and B.M. Friedrich \inst{2,3}}
\shortauthor{S. Lange \etal}

\institute{                    
  \inst{1} HTW, Dresden, Germany \\
  \inst{2} Center for Advancing Electronics Dresden, TU Dresden, Dresden, Germany \\
  \inst{3} Cluster of Excellence `Physics of Life', TU Dresden, Dresden, Germany
}

\abstract{%
We report an empirical power law for the reduction of network permeability 
in statistically homogeneous spatial networks upon removal of a single edge. 
We characterize this power law for plexus-like microvascular sinusoidal networks from liver tissue, 
as well as perturbed two- and three-dimensional regular lattices.
We provide a heuristic argument for the observed power law by mapping 
arbitrary spatial networks that satisfies Darcy's law
on an small-scale resistor network.
}

\begin{document}

\maketitle

% TITLE ==================================================================================

\date{\today}

\maketitle

% ----------------------------------------------------------------------------------------
\section{Introduction}

Spatial transport networks are found in biology, 
e.g., leaf venation networks \cite{Katifori2016}, fungal mycelium \cite{Tero2010,Alim2013},
and microvasculature \cite{Debbaut2012,Piergiovanni2017,Chang2017,Morales2019,Karschau2020}, 
as well as in technical systems, 
e.g., power grids \cite{Witthaut2016,Schafer2018}. % and human transportation networks
The resilience of such networks against failure of individual edges is thus both of practical and theoretical interest. 

Previous research addressed 
the trade-off between network resilience and the building and repair costs of networks \cite{Farr2014,Bottinelli2017}, 
as well as self-organized networks that optimize resilience and 
adaptation to fluctuations in load \cite{Katifori2016,Ronellenfitsch2019}.
Different measures were proposed to quantify network resilience, e.g.,
permeability-at-risk, i.e., the reduction of permeability as function of the fraction removed edges \cite{Karschau2020}, or 
the probability of percolation upon removal of a single edge \cite{Katifori2016}.
The later provides a link to the bond percolation problem in the theory of random resistor networks \cite{Kirkpatrick1973,Redner2011}.

Both DC electrical resistor networks and biological flow networks are well described by Kirchoff equations 
of current conservation and an effective Ohm's law \cite{Barthelemy2011}.
In the case of flow networks, this is a consequence of Poiseuille's law 
for Stokes flow in pipes at low Reynolds numbers,
which states a linear relation between the pressure difference at the end points of an edge and the current through that edge
\cite{Happel:book}. 

Previous work either solved the system of Kirchhoff equations for an entire network
\cite{Bottinelli2017,Katifori2016,Farr2014,Ronellenfitsch2019,Debbaut2012,Schliess2014,Schwen2014}, 
or resorted to continuum models of a homogenized effective medium.
For example, continuum models have been applied to model pharmacological relevant blood and bile flow in liver tissue 
\cite{Bonfiglio2010,Debbaut2012,Meyer2017,Mosharaf2019}.

Here, we report an empirical power law 
for the reduction of network permeability, 
which we observe in sinusoidal microvasculature networks from liver tissue,
as well as perturbed regular networks. 
We find that upon removal of a random edge, the permeability of these network is reduced 
by a relative amount that scales \textit{quadratically} with the current through that edge in the unperturbed network. 

We provide a heuristic explanation for this power law:
we map a spatial network with a distinguished central edge to 
an effective small-scale resistor network.
This allows us to relate the change in permeability upon removal of the central edge 
and the current through this central edge in the unperturbed network. 

% ----------------------------------------------------------------------------------------
\section{Permeability of spatial networks}

\begin{figure}
\includegraphics[width=\linewidth]{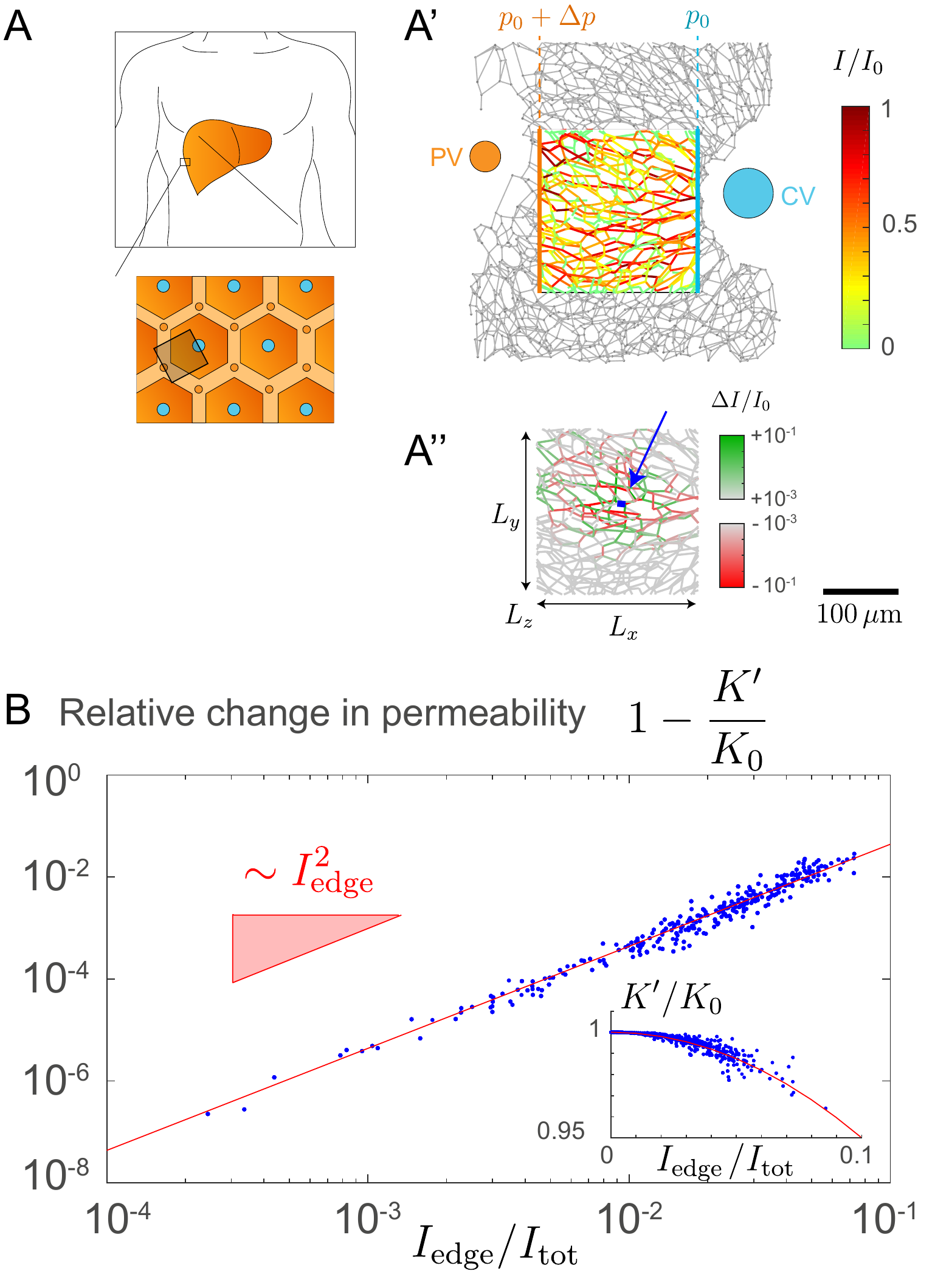}
\caption{
\textbf{Empirical power law for the reduction of network permeability for a sinusoidal network in liver tissue.}
\textbf{A}. Schematic representation of liver tissue with array of millimeter-sized liver lobules.
\textbf{A'}. Digital reconstruction of sinusoidal network in a single liver lobule.
Blood flows from the portal vein (PV, orange) to the central vein (CV, cyan). 
We computed flow patterns in a region of interest, using simplified boundary conditions of a pressure difference 
$\Delta p$ at opposite sides.
\textbf{A''}. Redistribution of computed flow upon removal of a single edge (blue, indicated by arrow).
Reference current $I_0=\Delta p/(\kappa L_x)$, scale bar: $100\,\mu\mathrm{m}$.
\textbf{B}. 
Relative change in permeability of the sinusoidal network upon removal of a single edge
as function of the current $\Iedge$ through that edge in the unperturbed network
($K_0$ permeability of unperturbed network, $K'$ permeability after removal of single edge, 
$\Itot$ total current through unperturbed network).
}
\label{figure1}
\end{figure}

We consider special cases of spatial networks with a characteristic mesh size $a$, 
inside a cuboid region of interest % (aligned with the axes of an $xyz$ coordinate system)
of dimensions $L_x \times L_y \times L_z$, see Fig.~\ref{figure1}A. % TODO

We impose the pressure $p_0+\Delta p$ at all network nodes at $x=0$ (source nodes), 
and the pressure $p_0$ at all network nodes at $x=L_x$ (sink nodes). 
An analogous problem formulation using voltages and electric currents is formally equivalent. 
We solve the Kirchhoff equations for transport within the network 
by imposing a current conservation equation at each interior node and an Ohm's law at each network edge.
The resistance $R_{ij}$ of an edge $(i,j)$ is proportional to its Euclidean length $l_{ij}$, 
$R_{i,j} = \kappa\, l_{ij}$, 
where $\kappa$ denotes a constant resistance per unit length.
(Alternatively, we may use the weight of the edge in the case of a weighted graph). 

We define the \textit{normalized permeability} $K_0$ of the network as \cite{Whitaker1986,Karschau2020}
\begin{equation}
K_0 = \frac{\kappa L_x}{A}\, \frac{I_\mathrm{tot}}{\Delta p},
\end{equation}
with units of an area density, 
where $I_\mathrm{tot}$ denotes the total current through the unperturbed network 
and $A=L_yL_z$ its cross-sectional area.
The normalization factor $\kappa L_x/A$ turns $K_0$ into a `material constant' of the network
that depends on the statistics of its local geometry, but is independent of its size.
This is essentially Darcy's law \cite{Whitaker1986}.
Darcy's law was originally formulated for flow through porous media, 
yet it applies also to statistically homogeneous spatial networks as a special case.
Note that for a network consisting of $n$ straight lines parallel to the $x$ axis 
that connect $x=0$ and $x=L$, we would have $K_0=n/A$.

We investigated the \textit{change} in network permeability upon removal of a single edge. 
Let $K'$ denote the permeability of the perturbed network and $\Iedge$ the current that flew through this edge before its removal.
We empirically observed an approximate power law with exponent two
\begin{equation}
\label{eq_powerlaw}
1 - \frac{K'}{K_0} \sim \Iedge^2 \quad
\end{equation}
with some factor of proportionality $\gamma$.
This empirical observation was made for different types of networks.
Eq.~(\ref{eq_powerlaw}) characterizes, e.g., flow computations for sinusoidal networks in liver tissue, 
see Fig.~\ref{figure1}.
Here, a nematic axis of network alignment \cite{Morales2019,Karschau2020} % TODO: add Scholich2019
is oriented parallel to the $x$ axis.
Analogous results were found for flow along the $y$ axis, i.e., perpendicular to the nematic axis of network alignment (not shown).
A linear regression for the relative change of permeability in sinusoidal networks 
predict a power law exponent with 95{\%}-confidence interval $[2.03,2.05]$ for the network shown in Fig.~\ref{figure1}
(similarly, $[1.99,2.00]$ and $[2.04,2.05]$ for two additional data sets from \cite{Karschau2020}).

\begin{figure}
\includegraphics[width=\linewidth]{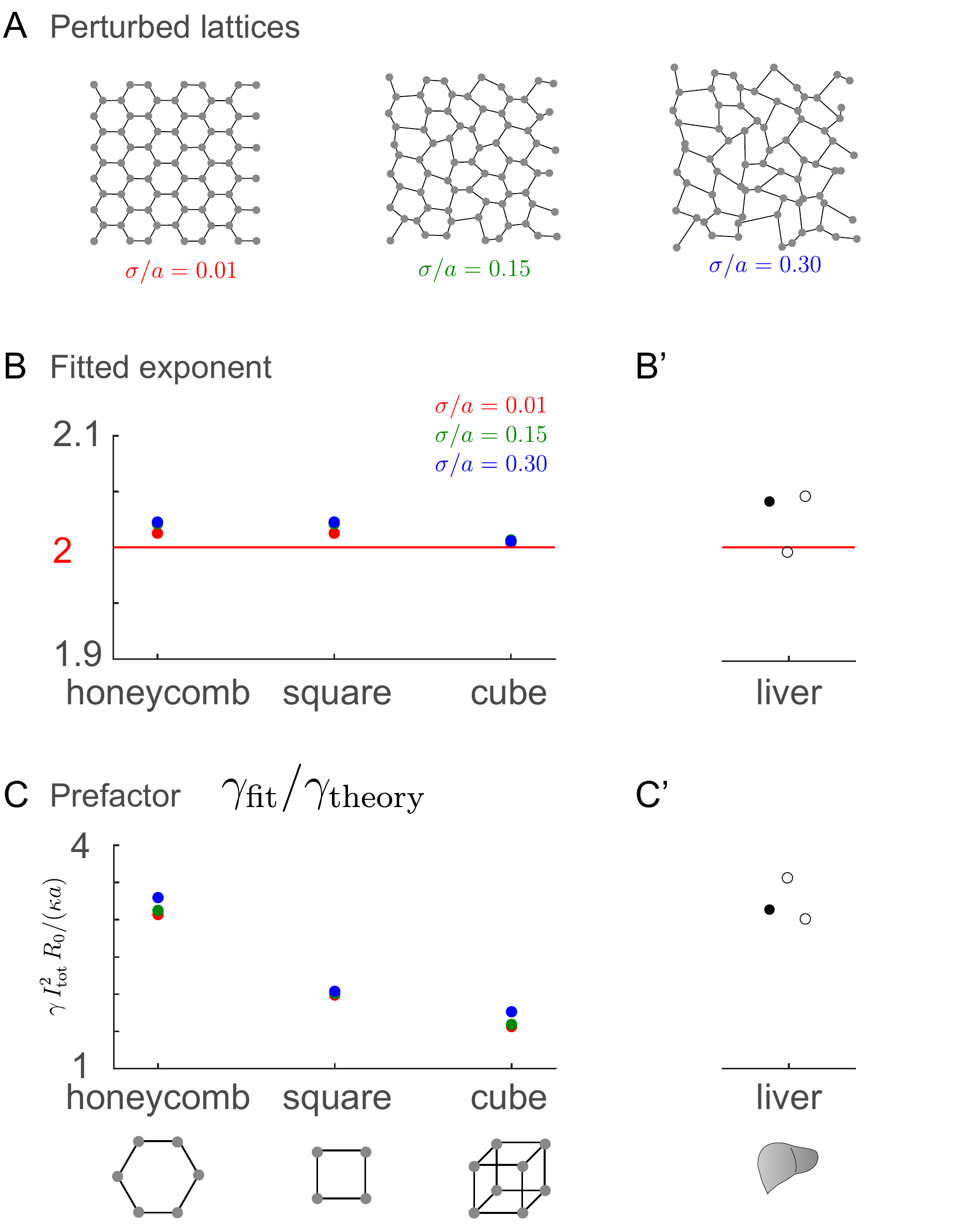}
\caption{
\textbf{Reduction of network permeability in perturbed regular lattices.}
\textbf{A}.
Sections of perturbed regular lattice (honeycomb lattice)
for three different perturbation strengths $\sigma/a$ 
(using normally distributed independent random perturbations of variance $\sigma^2$ added to each coordinate of each node).
\textbf{B}. 
Exponents $\beta$ of fitted power laws for the relative change in network permeability upon removal of a single edge
according to $1-K'/K_0 \sim \Iedge^\beta$:
all fitted exponents are close the theoretical prediction $\beta_\mathrm{theory} = 2$, see Eq.~(\ref{eq_powerlaw}).
Simulations were performed for perturbed honeycomb, square and cubic lattices,
for three different perturbation strengths $\sigma/a$ as shown.
% (using normally distributed independent random perturbations of variance $\sigma^2$ added to each coordinate of each node).
\textbf{C}. 
Fitted prefactor $\gamma_\mathrm{fit}$ of the power law Eq.~(\ref{eq_powerlaw}),
$1-K'/K_0=\gamma\,\Iedge^2$,
normalized by the theoretical order-of-magnitude estimate 
$\gamma\sim\gamma_\mathrm{theory}$ 
with 
$\gamma_\mathrm{theory}=(\kappa a)/(R_0\,\Itot^2)$
from
Eq.~(\ref{eq:gamma}).
Network dimensions are: 
honeycomb lattice $60\times 60$, 
square lattice $60\times 60$, 
cubic lattice $20\times 20\times 10$
(corresponding to 3600, 3600, 4000 nodes, respectively).
\textbf{B', C'}. 
Analogous statistics for sinusoidal networks from liver tissue 
(filled symbol: network from Fig.~\ref{figure1}, open symbols: two additional networks from \cite{Morales2019};
same vertical axes as B,C).
}
\label{figure2}
\end{figure}

Additionally, we observed the power law Eq.~(\ref{eq_powerlaw}) in perturbed regular lattices, 
using honeycomb, square, and cubic lattices as prototypical examples, see Fig.~\ref{figure2}.
For these regular networks, 
we added isotropic Gaussian noise to node positions.
(Using log-normally distributed random edge weights gave analogous results, not shown).
Interestingly, for these regular networks, 
any deviations from the power law Eq.~(\ref{eq_powerlaw}) always occurred only for special edges: 
these edges were either close to the boundary of the region of interest, or carried an unusual high current. 

% ----------------------------------------------------------------------------------------
\section{Origin of the power law}

We consider an edge connecting nodes labeled $1$ and $2$, 
carrying a directed current $\Iedge$ in the unperturbed network.
We will replace this edge by an equivalent source-sink dipole and map the flow problem 
on the small-scale resistor network shown in Fig.~\ref{figure3}.

Specifically, we can replace the link $(1,2)$ by 
a sink of strength $-\Iedge$ placed at node $1$, and
a source of strength $\Iedge$ placed at node $2$.
Note that inserting a source of strength $I$ at an interior node of the network
will decrease the inflow $I_\mathrm{in}$ at the left boundary $x=0$ by an amount $-\alpha I$,
and increase the outflow $I_\mathrm{out}$ at the right boundary $x=L$ by an amount $(1-\alpha)I$.
We expect $\alpha\approx (L-x)/L$, where $x$ is the coordinate of the node
along the direction of the pressure gradient 
(this holds exactly true, e.g., for a perfect square lattice). 
Obviously, if a sink is placed at node $1$ and a source at node $2$, 
both nodes might have different splitting ratios $\alpha_1$ and $\alpha_2$, respectively. 
This results in a net change of the total current through the network
upon insertion of the source-sink dipole (or equivalently insertion of the edge $(1,2)$)
\begin{equation}
\label{eq:Itot}
\Itot = \Itot' + (\alpha_1-\alpha_2)\Iedge \quad.
\end{equation}
Here, $\Itot$ is the current through the network 
with edge $(1,2)$ still present (or, equivalently, source-sink dipole added as described above), 
and $\Itot'$ is the current through the network without edge $(1,2)$.
The important point is now that the difference $\alpha_1-\alpha_2$ and the current $\Iedge$ are highly correlated. 

% ----------------------------------------------------------------------------------------
\subsection{Equivalent small-scale resistor network}
\begin{figure}
\includegraphics[width=\linewidth]{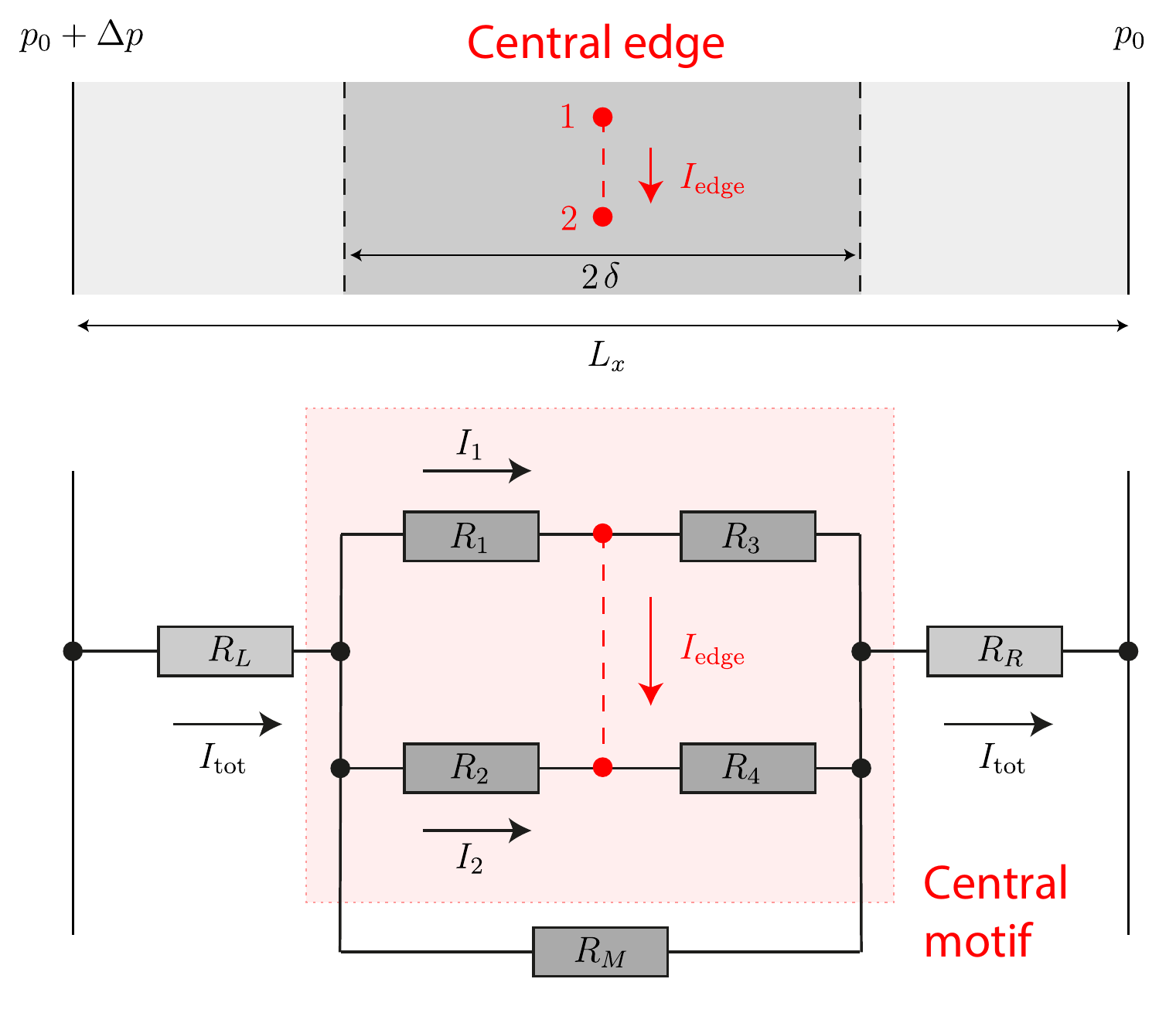}
\caption{
\textbf{
Re-routing in spatial networks upon removal of a single edge: 
mapping on small-scale resistor network.}
We consider a central edge of a homogeneous network that connects nodes $1$ and $2$ with directed current $\Iedge$. 
To estimate the relative change in network permeability upon removal of edge $(1,2)$, 
and map the full network on a small-scale resistor diagram, see text for details.
In particular, 
we introduce virtual boundaries at a distance $d$ from the central edge parallel to the boundaries of the network.
We consider effective resistances $R_L$ and $R_R$ for flow from the left and the right outer boundary to these virtual boundaries, 
as well as effective resistances $R_1$, $R_2$, $R_3$, $R_4$ for flow node $1$ and $2$ to the two virtual boundaries, respectively.
}
\label{figure3}
\end{figure}

To proceed, we replace the large network by a simple effective resistor network, see Fig.~\ref{figure3}.
Specifically, let $x_1$ and $x_2$ be the $x$ coordinates of node $1$ and $2$, and $x'=(x_1+x_2)/2$ their mean.
We expect that different splitting ratios $\alpha_1$ and $\alpha_2$ for nodes 1 and 2
reflect the local geometry of the network.
In contrast, we anticipate that on larger spatial scales $\delta$ with $a\ll \delta\ll L_x$, 
flow can be considered homogeneous and described by effective resistances, 
thus corroborating Darcy's law \cite{Whitaker1986}.
Specifically, we consider a plane normal to the $x$ axis placed at $x=x'-\delta$, 
and introduce an effective resistance $R_1$ for flow from node $1$ to this virtual plane. 
Likewise, we introduce an effective resistance $R_L$ for flow from this virtual plane to the left boundary at $x=0$.
% (As a technical point, the resistance $R_L$ should be determined for a distribution of sources at $x'-d$, corresponding to the distribution of currents through this boundary if a single source is placed at node $1$.)
Analogous to $R_1$ and $R_L$, 
we introduce a local resistance $R_3$ for flow from node $1$ to a second virtual plane at $x'+\delta$, 
as well as a resistance $R_R$ for flow from this plane to the right boundary at $x=L_x$.
For node $2$, we introduce resistances $R_2$ and $R_4$ analogous to $R_1$ and $R_3$, respectively, see Fig.~\ref{figure3}.
Finally, we introduce a resistance $R_M$ to account for flow 
between the two virtual boundaries far from the edge $(1,2)$.
% which is chosen such that the total resistance of the network is $R_0 = \kappa L_x/(K_0 A)$.

The length of edge $(1,2)$ is on the order of the mesh size $a$ of the network, 
and thus much smaller than the coarse-graining length-scale $\delta$.
We thus expect that the resistance $R_{12}$ of edge $(1,2)$ is much smaller than
the resistances $R_1$, $R_2$, $R_3$, $R_4$ of the central network motif. 
For the following calculation, we make the simplifying assumption that $R_{12}=0$.  
The current $\Iedge$ through the edge $(1,2)$ is found to be proportional to an imbalance of resistances, as expected
\begin{equation}
\label{eq:I12}
\Iedge 
= 
\frac{R_2 R_3 - R_1 R_4}{(R_1+R_2)(R_3+R_4)} (I_1 + I_2)
% \sim \frac{R_1}{R_3} - \frac{R_2}{R_4} 
% \sim \frac{R_1}{R_2} - \frac{R_3}{R_4} 
\quad,
\end{equation}
which implies that $\Iedge$ vanishes if $R_1:R_3 = R_2:R_4$.

The splitting ratios introduced above approximately satisfy
\begin{align}
\alpha_1 : 1-\alpha_1 &= R_1 + R_L:R_3+R_R \text{, and } \\
\alpha_2 : 1-\alpha_2 &= R_2 + R_L:R_4+R_R \quad.
\end{align}
In the limit $R_1, R_2, R_3, R_4\gg R_L, R_R$, 
this implies for their difference
\begin{equation}
\alpha_1 - \alpha_2 = \frac{R_2 R_3 - R_1 R_4}{(R_1+R_3)(R_2+R_4)} \sim \Iedge \quad.
\end{equation}
Together with Eq.~(\ref{eq:Itot}), 
Eq.~(\ref{eq_powerlaw}) follows.

A direct calculation provides an estimate for the proportionality factor $\gamma$ in Eq.~(\ref{eq_powerlaw})
(up to a network-type specific factor of order unity), see appendix
\begin{equation}
\label{eq:gamma}
\gamma \sim 
\gamma_\mathrm{theory} 
= \frac{a}{L_x} \frac{1}{I_0\, \Itot} 
= \frac{A\,a}{L_x}\, K_0 \, \Itot^{-2} 
= \frac{\kappa a}{R_0}\, \Itot^{-2} 
\quad, 
\end{equation}
where $I_0 = \Delta p/(\kappa L_x) = \Itot / (K_0 A)$.
In short, we can use Kirchhoff's laws to 
first compute the relative change in permeability upon removal of edge $(1,2)$ of the central network motif shown in Fig.~\ref{figure2}.
The hierarchy of length scales
between the mesh size $a$ of the network, the coarse-graining length scale $\delta$, and the total size $L_x$ of the network, $a\ll \delta\ll L_x$, 
implies a hierarchy of resistances according to Darcy's law 
with 
$R_1,R_2,R_3,R_4 \sim (\delta/\delta^2) \, \kappa/K_0$, % $i=1,\ldots,4$,
$R_M \sim (2\delta/A) \, \kappa/K_0$, 
$R_L+R_R \sim (L_x/A) \, \kappa/K_0$;
hence,
$R_M\ll R_1,R_2,R_3,R_4$ and $R_M\ll R_L+R_R$.
Exploiting this hierarchy of resistances allows us to relate 
the relative change in permeability of the central network motif 
to that of the full network.
According to Eq.~(\ref{eq:gamma}), 
$\gamma$ scales with the inverse square of the total current $I_\mathrm{tot}$ in the unperturbed network,
multiplied with a dimensionless factor given by the ratio of the average resistance $\kappa a$ of a single edge 
divided by the resistance $K_0$ of the full network.
Simulation results for perturbed regular lattices are consistent with this result, Eq.~(\ref{eq:gamma}), see Fig.~\ref{figure2}C.

For the sinusoidal network shown in Fig.~\ref{figure1}, 
we can define a proxy for the mesh size as the median of edge lengths in the network
$a=16.2\,\micron$
(mean$\pm$s.e.: $16.2\pm 10.4\,\micron$).
Using this value for $a$, we find for the ratio 
between the fitted factor of proportionality $\gamma_\mathrm{fit}$
from a fit of Eq.~(\ref{eq_powerlaw}) 
and the theoretical estimate $\gamma_\mathrm{theory}$ in Eq.~(\ref{eq:gamma})
$\gamma_\mathrm{fit}/\gamma_\mathrm{theory} \approx 3.1$
(similarly, we find 3.0 and 3.5 for two additional data sets from \cite{Karschau2020}).

Interestingly, $\gamma_\mathrm{fit}/\gamma_\mathrm{theory}$ approximately scales as $1/d$ in these examples, 
where $d$ is the degree of the network
(ho\-ney\-comb lattice: $d=3$, square lattice $d=4$, cubic lattice $d=6$, sinusoidal networks $d=3.3\pm 0.6$, mean$\pm$s.e.).

% ----------------------------------------------------------------------------------------
\section{Discussion}
% TODO: expand

We reported an empirical scaling law for the relative change in network permeability 
of homogeneous spatial transport networks with characteristic mesh-size, Eq.~(\ref{eq_powerlaw}),
testing both
sinusoidal blood networks in liver tissue as an example of a biological network, see Fig.~\ref{figure1}, as well as
perturbed regular lattices, see Fig.~\ref{figure2}.
We rationalize this empirical observation by mapping a generic homogeneous spatial network
on an effective small-scale resistor network, see Fig.~\ref{figure3}.

A symmetry argument already dictates the \textit{form} of the power law:
as the directed current $\Iedge$ of a central edge can have either sign, 
but the permeability must decrease upon removal of that edge 
as a consequence of Helmholtz' theorem of low-Reynolds number flows \cite{Happel:book},
the relative change of permeability cannot depend linearly on $\Iedge$, but quadratically at most. 
Similarly, dimensionality arguments constrain the possibilities for the factor of proportionality $\gamma_0$ in the power law.
Thus, it is not the form of the power law that is surprising, but the fact that such a power law exists in the first place.

Our finding highlights the importance of high-current edges for the resilience of transport networks against perturbations \cite{Karschau2020}.
Future work will address generalizations to time-varying networks \cite{Karschau2018}.

% -----------------------------------------------------------------------------------------------------------------------------
\acknowledgments
The authors are supported by the DFG through the Excellence Initiative 
by the German Federal Government and State Government:
Clusters of Excellence \textit{cfaed} (EXC 1056) and \textit{Physics of Life} (EXC 2068).
We thank 
% Lutz Brusch, 
Szabolcs Horv{\'a}t, 
Yannis Kalaidzidis, 
Felix Kramer,
% Hernan Morales-Navarette, 
% Kirstin Meyer,
Carl Modes, 
Malte Schr{\"o}der, 
% Fabian Segovia-Miranda, 
Marc Timme, 
Marino Zerial, 
as well as all members of the Biological Algorithms group for stimulating discussions.

% -----------------------------------------------------------------------------------------------------------------------------
\section{Methods} Data acquisition for sinusoidal networks.
As described previously \cite{Morales2015,Morales2016,Morales2019},
fixed tissue samples of murine liver were optically cleared and treated with fluorescent antibodies for fibronectin and laminin,
thus staining the extracellular matrix surrounding the sinusoids.
Subsequently, samples were imaged at high-resolution using multiphoton laser-scanning microscopy.
Three-dimensional image data was segmented and network skeletons computed using
MotionTracking image analysis software \cite{Morales2016}. 
The data sets analyzed in this study correspond to the same used in \cite{Morales2019}.

\section{Appendix}

Let $S_0$ and $S'$ be the resistances of the central motif in Fig.~\ref{figure3} with and without the central edge $(1,2)$, respectively. 
A direct calculation yields
\begin{align}
S_0 &= (R_1\oplus R_2) + (R_3\oplus R_4) \notag \\
    &= \frac{ R_1 R_2 R_3 + R_1 R_2 R_4 + R_1 R_3 R_4 + R_2 R_3 R_4 }{(R_1+R_2)(R_3+R_4)} \quad, \\
S'  &= (R_1+R_3) \oplus (R_2+R_4) \notag \\
    &= \frac{ (R_1+R_3)(R_2+R_4) }{ R_1+R_2+R_3+R_4 } \quad.
\end{align}
Here, we used notation $a\oplus b = (a^{-1} + b^{-1})^{-1}$ for the effective resistance of two parallel resistors $a$ and $b$.
(Note that $\oplus$ is commutative and that the associate law holds, while the distributive law does not.)

Hence,
\begin{equation}
\label{eq:S0Sp}
1-\frac{S_0}{S'} = \frac{ (R_2 R_3 - R_1 R_4)^2 }{ (R_1+R_2)(R_1+R_3)(R_2+R_4)(R_3+R_4) } \quad.
\end{equation}
For the resistances of the full resistor diagram, we have
\begin{align}
R_0 &= R_L + (S_0\oplus R_M) + R_R \quad, \\
R'  &= R_L + (S' \oplus R_M) + R_R \quad.
\end{align}
In the main text, we argue that the hierarchy of length scales, $a\ll \delta\ll L$, 
implies
$R_M\ll R_1,R_2,R_3,R_4$ and $R_M\ll R_L+R_R$
as a consequence of Darcy's law 
with 
$R_i \sim (\delta/\delta^2) \, \kappa/K_0$, $i=1,\ldots,4$, as well as
$R_M \sim (2\delta/A) \, \kappa/K_0$, and
$R_L+R_R \sim (L_x/A) \, \kappa/K_0$.
Thus, $S_0, S'\gg R_M$, as well as $R_L, R_R \gg R_M$.
With this approximation,  
we find for the relative change in permeability upon removal of edge $(1,2)$ for the full network
\begin{align}
1-\frac{K'}{K_0} &= 1-\frac{R_0}{R'} \notag \\
&= 
1- \frac{ R_L + (S_0\oplus R_M) + R_R }{ R_L + (S' \oplus R_M) + R_R } \notag \\
&\approx \frac{1}{R_L+R_R} \left( (S'\oplus R_M) - (S\oplus R_M) \right) \\
&\approx \frac{R_M^2}{R_L+R_R} \left( \frac{1}{S_0} - \frac{1}{S'} \right) \\
&= \frac{1}{S_0}\,\frac{R_M^2}{R_L+R_R} \left( 1 - \frac{S_0}{S'} \right) 
\quad.
\label{eq:K0Kp}
\end{align}
The current $I_1+I_2$ through the central motif is
\begin{equation}
I_1 + I_2 
=       \frac{R_M}{R_M + S_0 } \Itot 
\approx \frac{R_M}{      S_0 } \Itot 
\quad;
\end{equation}
hence Eq.~(\ref{eq:I12}) yields
\begin{equation}
\Iedge = \frac{R_2 R_3 - R_1 R_4}{(R_1+R_2)(R_3+R_4)}\,\frac{ R_M }{ S_0 } \Itot \quad.
\end{equation}
We can thus rewrite Eq.~(\ref{eq:S0Sp}) as
\begin{align}
1-\frac{S_0}{S'} 
% \frac{ (R_1+R_2)^2(R_3+R_4)^2 }{ (R_1+R_2)(R_1+R_3)(R_2+R_4)(R_3+R_4) } 
% \left( \frac{ R_M + S_0}{R_M} \right)^2 \left( \frac{\Iedge}{I_0} \right)^2
&= 
\frac{ (R_1+R_2)(R_3+R_4) }{ (R_1+R_3)(R_2+R_4) } 
\left( \frac{S_0}{R_M} \right)^2 
\left( \frac{\Iedge}{\Itot} \right)^2 \notag \\
&\approx 
\left( \frac{ S_0}{R_M} \right)^2 
\left( \frac{\Iedge}{\Itot} \right)^2 
\quad.
\end{align}
Here, we used in the last step that $R_1$, $R_2$, $R_3$, $R_4$ will be approximately of equal magnitude.
Inserting the last result into Eq.~(\ref{eq:K0Kp}) yields
\begin{align}
1-\frac{K'}{K_0}
&\approx \frac{ S_0 }{ R_L+R_R } \left( \frac{\Iedge}{\Itot} \right)^2 \notag \\
&\sim \frac{\kappa a}{R_0} \left( \frac{\Iedge}{\Itot} \right)^2 \notag \\
% &= \frac{L_y L_z}{L_x} a K_0 \left( \frac{\Iedge}{\Itot} \right)^2 \\
&= \frac{a}{L_x} \frac{\Iedge^2}{I_0\, \Itot} 
\quad.
\end{align}
Here, we used that the coarse-graining distance $\delta$ should be chosen larger, 
but proportional (and on the same order of magnitude) as the mesh size $a$ of the network;
thus, we expect that $S_0$ scales proportional to $\kappa a$.
Additionally, since $\delta \ll L_x$, we have $R_L + R_R\approx R_0$.

\bibliography{lange_networks_powerlaw}

\end{document}